\documentstyle[12pt]{article}
\def \apk{A_{\pi K}}
\def \abpk{\overline{A}_{\pi K}}
\def \app{A_{\pi \pi}}
\def \abpp{\overline{A}_{\pi \pi}}
\def \beq{\begin{equation}}
\def \cap{{\cal P}}

\def \cat{{\cal T}}

\def \eeq{\end{equation}}
\def \ob{\overline{B}}
\def \ok{\overline{K}}

\def \tcapp{\tilde{\cal P'}}
\def \tru{\tilde r_u}
\begin{document}
\renewcommand{\thetable}{\Roman{table}}
\rightline{CERN-TH/96-147}
\rightline{EFI-96-20}
\rightline{hep-ph/9606207}
\rightline{June 1996}
\bigskip
\bigskip
\centerline{\bf DISCRETE AMBIGUITIES}
\centerline{{\bf IN EXTRACTING WEAK PHASES FROM $B$ DECAYS}
\footnote{To be submitted to Phys.~Rev.~D.}}
\bigskip
\centerline{\it Amol S. Dighe}
\centerline{\it Enrico Fermi Institute and Department of Physics}
\centerline{\it University of Chicago, Chicago, IL 60637}
\medskip
\centerline{and}
\medskip
\centerline{\it Jonathan L. Rosner}
\medskip
\centerline{\it Div.~TH, CERN}
\centerline{\it 1211 CH Geneva 23, Switzerland}
\smallskip
\centerline{and}
\smallskip
\centerline{\it Enrico Fermi Institute and Department of Physics}
\centerline{\it University of Chicago, Chicago, IL 60637
\footnote{Permanent address.}}
\bigskip
\bigskip
\bigskip
\bigskip
\centerline{\bf ABSTRACT}
\medskip
\begin{quote}

Phases of elements of the Cabibbo-Kobayashi-Maskawa (CKM) matrix, as obtained
using decays of $B$ mesons to $\pi^+ \pi^-$, $\pi^\pm K^\mp$, and $\pi^+ K^0$
or $\pi^- \ok^0$, are shown to have a class of discrete ambiguities. In most
cases these can be eliminated using other information on CKM phases. 
\end{quote}

\leftline{PACS codes:  11.30.Er, 12.15.Hh, 13.25.Hw, 14.40.Nd}
\vfill
\leftline{CERN-TH/96-68}
\leftline{June 1996}
\newpage

A promising source of information about the mechanism of CP violation is the
study of rate asymmetries in the comparison of $B$ and $\ob$ decays to specific
final states.  These asymmetries often involve unknown strong-interaction phase
shifts.  A method was recently proposed \cite{DGR} to circumvent this
difficulty using time-dependent $B^0$ and $\ob^0$ decays and time-integrated
rates for $B^0 \to \pi^- K^+$, $\ob^0 \to \pi^+ K^-$, and $B^+ \to \pi^+ K^0$
or $B^- \to \pi^- \ok^0$.  (The last two rates are predicted to be equal.)
Within an assumption of flavor SU(3) for strong phase shifts and for diagrams
dominated by tree (but not penguin) graphs, it was possible to exhibit six
equations in six unknowns and thus to demonstrate the existence of solutions
for all parameters of interest.  However, the Monte Carlo method employed in
Ref.~\cite{DGR} indicated the presence of discrete ambiguities.  Using
numerical methods in the present note, we clarify these ambiguities, and show
that they may be eliminated for the most part using other information already
known about phases of the Cabibbo-Kobayashi-Maskawa (CKM) matrix. 

The amplitudes of the processes $B^0\to\pi^+\pi^-$ and $B^0\to\pi^- K^+$ are
defined as $\app$ and $\apk$, while that for $B^+\to \pi^+ K^0$ is defined as
$A_+$. The amplitudes for the corresponding charge-conjugate decay processes
are denoted by $\abpp,~\abpk,~A_-$, respectively. It was shown in
Ref.~\cite{DGR} that one can measure six independent combinations of the
following six parameters:  the strangeness-preserving tree amplitude $\cat$,
the strangeness-preserving and -violating penguin amplitudes $\cap$ and
$\tcapp$, the weak phases $\alpha$ and $\gamma$, and the strong phase $\delta$.
These combinations may be expressed as 

\beq \label{A}
A \equiv \frac{1}{2}(|\app|^2 + |\abpp|^2)
= \cat^2 + \cap^2 - 2 \cat \cap \cos \delta \cos \alpha~~~,
\eeq
\beq \label{B}
B \equiv \frac{1}{2}(|\app|^2 - |\abpp|^2)
= - 2 \cat \cap \sin \delta \sin \alpha~~~,
\eeq
\beq \label{C}
C \equiv {\rm Im}~(e^{2 i \beta} \app \abpp^*)
= - \cat^2 \sin 2\alpha + 2 \cat \cap \cos \delta \sin \alpha~~~,
\eeq
\beq \label{D}
D \equiv \frac{1}{2}(|\apk|^2 + |\abpk|^2)
= (\tru \cat)^2 + \tcapp^2 - 2 \tru \cat \tcapp \cos \delta \cos \gamma~~~~,
\eeq
\beq \label{E}
E \equiv \frac{1}{2}(|\apk|^2 - |\abpk|^2)
= 2 \tru \cat \tcapp \sin \delta \sin \gamma~~~,
\eeq
\beq \label{F}
F \equiv |A_+|^2 = |A_-|^2 = \tcapp^2~~~.
\eeq

Here $\tru \equiv r_u f_K/f_\pi$, where $r_u \equiv |V_{us}/V_{ud}| = 0.23$.
The quantities $A - C$ are measured in time-dependent rates for $B^0$ or
$\ob^0 \to \pi \pi$, $D$ and $E$ by comparing rates for $B^+ \to \pi^- K^+$ and
$B^- \to \pi^+ K^-$, and $F$ via the rate for the process $B^+ \to \pi^+ K^0$,
which is predicted to be dominated by a single penguin amplitude and hence to
have the same rate as $B^- \to \pi^- \ok^0$. 

We considered \cite{DGR} a set of representative CKM elements parametrized
\cite{WP} as shown in Table I, where $\rho$ and $\eta$ are the real and
imaginary parts of $V^*_{ub}/|V_{cd}V_{cb}|$.  For each of these points,
the phase shifts $\delta = 5.7^\circ$, $36.9^\circ$, $84.3^\circ$,
$95.7^\circ$, $143.1^\circ$, and $174.3^\circ$ were chosen.  (We shall not be
concerned here with $\sin \delta = 0$, a singular case in which the above
equations no longer provide sufficient information.)  Monte Carlo results
indicated that the equations sometimes had more than one solution.

\begin{table}
\caption{Points in the $(\rho,\eta)$ plane and angles of the unitarity
triangle.} 
\begin{center}
\begin{tabular}{c c c c c c} \hline
Point & $\rho$ & $\eta$ & $\alpha$ & $\beta$ & $\gamma$ \\
      &        &        &  (deg.)  &  (deg.) &  (deg.)  \\ \hline
$p_1$ &$-0.30$ &  0.15  &   20.0   &   6.6   &  153.3   \\ 
$p_2$ &   0    &  0.35  &   70.7   &  19.3   &   90.0   \\
$p_3$ &  0.36  &  0.27  &  120.3   &  22.9   &   36.9   \\ \hline
\end{tabular}
\end{center}
\end{table}

We have used an exact numerical method to obtain all solutions of
Eqs.~(\ref{A}-\ref{F}) for the points $p_1$, $p_2$, $p_3$ and the six phases
$\delta$. We express the five observables $A, B, C, D, E$ in terms of five
unknowns $\cat, \cap, \alpha, \gamma, \delta$ by substituting the measured
value of $\tcapp = \sqrt{F}$ and noting that $\tru$ also is well-measured. The
solution then proceeds as follows: 

\begin{itemize}
\item We eliminate $\gamma$ from Eq.~(\ref{D}) and Eq.~(\ref{E}) to get
\beq
D = \tru ^2 \cat ^2 + F - 2 \tru \cat \sqrt{F} \cos \delta \left( 1 -
	\frac{E^2}{4 \tru ^2 \cat ^2 F \sin ^2 \delta} \right) ^{1/2}~~~.
\eeq
When both the sides are squared, this equation becomes a quadratic in $x \equiv
\sin ^2 \delta$ whose coefficients depend only on $\cat$. For each of the
solutions which is real and lies between 0.0 and 1.0 (since $x = \sin ^2
\delta$), 

\item we eliminate $\alpha$ from Eq.~(\ref{A}) and Eq.~(\ref{B}) to get
\beq
A = \cat ^2 + \cap ^2 - 2 \cat \cap \cos \delta \left( 1 - 
	\frac{B^2}{4 \cat^2 \cap^2 \sin ^2 \delta} \right) ^{1/2}~~~.
\eeq
When both the sides of this equation are squared, we obtain a quadratic in
$y \equiv \cap^2$ whose coefficients involve only $\cat$ and $x$, which is a 
known function of $\cat$. We proceed with those values of $y$ that are real and
positive.

\item Now we know all the other unknowns $\cap, \alpha, \gamma, \delta$ as
explicit functions of a single unknown $\cat$. We can now check for those
values of $\cat$ which satisfy Eq.(\ref{C}). We can increase the accuracy of
our solutions as much as we want by decreasing the step size in $\cat$  and
using the {\it zero crossing algorithm}, where a solution corresponds to that
value of $\cat$ where increasing $\cat$ by a small amount changes the sign of
[L.H.S. - R.H.S.] in Eq.~(\ref{C}). 

\end{itemize}

As many as 8 solutions were found for some sets of input parameters. The
results are summarized in Tables II -- IV for points $p_1 - p_3$.  We calculate
$A - E$ for the input values $\cat = 1$, $\tcapp = 1$, $\cap = \tcapp r_u \sin
\gamma/\sin \alpha$ [assuming flavor SU(3) for the input], and the input strong
phases shown in the Tables. The equations are then inverted using the method
described above to obtain the output phases.  In some cases the numerical
algorithm gives two closely related or identical sets of output phases; we have
indicated these with equal numbers. These are probably identical solutions
arrived at through two different branches of the step-by-step method described
above, with small differences associated with rounding errors.  Nonetheless, we
feel this point could benefit from further study. 

\begin{table} 
\caption{Output values of weak and strong phases, for given values of
input strong phases, in degrees, for the point $p_1$ with $\alpha_{\rm in}
= 20.0^\circ$ and $\gamma_{\rm in} = 153.3^\circ$.}
\begin{center}
\begin{tabular}{r r r r l} \\ \hline \hline
$\delta_{\rm in}$ & $\alpha_{\rm out}$ & $\gamma_{\rm out}$ &
$\delta_{\rm out}$ & Notes \\ \hline \hline
  5.7 &  20.0 & 153.4 &   5.7 & (a) \\
      &  10.4 & 106.1 &   1.4 & (b) \\ \hline
 36.9 &  20.0 & 153.4 &  36.9 & (a) \\ \hline
 84.3 &  20.0 & 153.4 &  84.3 & (a) \\
      &  70.6 & 153.8 &  84.6 & (c) \\
      &  21.5 &  28.7 &  97.5 & (b) \\
      &  71.8 &  26.4 &  95.5 & (b) \\
      &  59.3 &  93.1 &  23.6 & (d) \\
      &  29.9 &  70.8 & 136.5 & (b) \\
      &  82.8 &  83.4 & 152.3 & (d) \\ \hline
 95.7 &  18.8 &  25.0 &  83.6 & (b) \\
      &  71.5 &  24.8 &  83.5 & (b) \\
      &  20.0 & 153.4 &  95.7 & (a) \\
      &  72.8 & 155.0 &  96.4 & (c) \\
      &  60.8 &  82.6 &  22.7 & (d) \\
      &  29.3 &  91.9 & 140.3 & (b) \\
      &  83.2 &  96.0 & 154.0 & (e) \\ \hline
143.1 &  72.2 &  16.6 &  48.6 & (b) \\
      &  14.4 &  14.6 &  50.8 & (b) \\
      &  20.0 & 153.2 & 143.3 & (a,1) \\
      &  20.0 & 153.4 & 143.1 & (a,1) \\
      &  76.3 & 162.9 & 132.0 & (c) \\
      &  66.4 &  49.6 &  15.7 & (b) \\
      &  81.7 & 132.6 & 162.4 & (c) \\ \hline
174.3 &  74.1 &   3.1 &  40.2 & (b) \\
      &  15.0 &   3.0 &  41.1 & (b) \\
      &  15.9 & 176.7 & 140.9 & (c) \\
      &  75.0 & 176.9 & 139.9 & (b) \\
      &  20.0 & 153.4 & 174.3 & (a) \\
      &  81.1 & 140.7 & 176.8 & (c) \\ \hline \hline
\end{tabular}
\end{center}
\leftline{(a) Correct solution; (b) $\beta > \pi/4$ ; (c) $\alpha + \gamma
> \pi$;}
\leftline{(d) potential ambiguity; (e) $\beta$ or $\gamma$ too small.}
\leftline{Numbers denote solutions probably identical to one another.}
\end{table}
\bigskip

\begin{table}
\caption{Output values of weak and strong phases, for given values of
input strong phases, in degrees, for the point $p_2$ with $\alpha_{\rm in}
= 70.7^\circ$ and $\gamma_{\rm in} = 90.0^\circ$.  Notes are as for Table II.}
\begin{center}
\begin{tabular}{r r r r l} \\ \hline \hline
$\delta_{\rm in}$ & $\alpha_{\rm out}$ & $\gamma_{\rm out}$ &
$\delta_{\rm out}$ & Notes \\ \hline \hline
  5.7 &   5.7 & 173.8 &  88.7 & (e) \\
      &  84.6 & 173.9 &  89.1 & (c) \\
      &   5.8 &   6.3 &  91.6 & (b) \\
      &  70.7 &  90.0 &   5.7 & (a) \\
      & 171.5 &  82.0 & 170.7 & (c) \\
      &  17.8 &  62.9 & 158.1 & (b) \\
      &  98.6 &  89.2 & 174.0 & (c) \\ \hline
 36.9 &  15.2 & 127.2 &  82.6 & (d) \\
      &  82.0 &  39.0 &  91.0 & (b) \\
      &  70.7 &  90.0 &  36.9 & (a) \\
      &  92.5 &  88.8 & 140.6 & (c) \\ \hline
 84.3 &  71.1 &  90.7 &  86.3 & (a,1) \\
      &  70.7 &  90.0 &  84.3 & (a,1) \\ \hline
 95.7 &  69.2 &  80.2 &  89.1 & (d) \\
      &  69.5 &  98.9 &  90.8 & (d) \\
      &  24.2 &  87.5 &  71.3 & (b) \\
      &  65.8 &  88.0 &  74.2 & (d) \\
      &  19.8 &  83.7 &  52.1 & (b) \\
      &  70.7 &  90.0 &  95.7 & (a) \\ \hline
143.1 &  30.8 &  34.3 &  88.8 & (b) \\
      &  61.3 &  33.7 &  88.5 & (b) \\
      &  31.1 & 145.4 &  91.1 & (d) \\
      &  61.0 & 146.2 &  91.5 & (c) \\
      &  50.2 &  86.2 &  29.3 & (d) \\
      &  19.5 &  81.1 &  21.8 & (b) \\
      &  39.8 &  86.4 & 135.2 & (b) \\
      &  70.7 &  90.0 & 143.1 & (a) \\ \hline
174.3 &  31.4 & 174.8 &  91.6 & (c) \\
      &  58.9 & 174.8 &  91.6 & (c) \\
      &  46.8 &  85.8 &   4.4 & (b) \\
      &  19.5 &  80.8 &   3.4 & (b) \\
      &  43.3 &  87.1 & 173.1 & (b) \\
      &  70.7 &  90.0 & 174.3 & (a) \\ \hline \hline
\end{tabular}
\end{center}
\end{table}
\bigskip

\begin{table}
\caption{Output values of weak and strong phases, for given values of
input strong phases, in degrees, for the point $p_1$ with $\alpha_{\rm in}
= 120.3^\circ$ and $\gamma_{\rm in} = 36.9^\circ$.  Notes are as for Table II.}
\begin{center}
\begin{tabular}{r r r r l} \\ \hline \hline
$\delta_{\rm in}$ & $\alpha_{\rm out}$ & $\gamma_{\rm out}$ &
$\delta_{\rm out}$ & Notes \\ \hline \hline
  5.7 & 141.6 &   5.0 &  40.0 & (e) \\
      & 126.8 &   4.9 &  40.3 & (b) \\
      & 128.4 & 175.2 & 138.8 & (c) \\
      & 143.2 & 175.3 & 138.6 & (c) \\
      & 135.7 &  34.0 &   6.5 & (d) \\
      & 120.3 &  36.9 &   5.7 & (a) \\
      & 136.2 & 134.6 & 176.1 & (c) \\
      & 151.4 & 131.9 & 176.7 & (c) \\ \hline
 36.9 & 120.4 &  36.8 &  37.0 & (a,1) \\
      & 120.9 &  34.2 &  39.5 & (a,1) \\
      & 129.8 & 154.5 & 131.9 & (c) \\
      & 146.4 & 155.8 & 130.6 & (c) \\
      & 120.2 &  37.2 &  36.8 & (a,1) \\
      & 135.2 & 128.8 & 157.9 & (c) \\
      & 152.7 & 125.7 & 161.5 & (c) \\ \hline
 84.3 & 145.8 &  38.2 &  84.6 & (c) \\
      & 120.3 &  36.9 &  84.3 & (a) \\
      & 147.4 & 143.6 &  95.9 & (c) \\
      & 121.9 & 143.8 &  95.9 & (c) \\
      & 137.6 &  86.7 &  47.8 & (c) \\
      & 111.1 &  85.1 &  40.4 & (c) \\
      & 132.3 &  97.2 & 148.9 & (c) \\
      & 158.6 & 100.7 & 156.0 & (c) \\ \hline
 95.7 & 146.8 & 140.0 &  82.3 & (c) \\
      & 118.6 & 142.4 &  83.8 & (c) \\
      & 148.8 &  37.2 &  96.0 & (c) \\
      & 120.3 &  36.9 &  95.7 & (a) \\
      & 139.7 & 105.3 &  51.2 & (c) \\
      & 109.6 &  98.6 &  41.2 & (c) \\
      & 130.9 &  87.8 & 148.5 & (c) \\
      & 160.5 &  94.6 & 157.2 & (c) \\ \hline
143.1 & 120.3 &  36.9 & 143.1 & (a,1) \\
      & 158.9 &  25.4 & 129.1 & (c) \\
      & 120.3 &  36.9 & 143.1 & (a,1) \\
      & 120.2 &  37.0 & 143.3 & (a,1) \\
      & 121.7 &  42.6 & 148.4 & (d) \\ \hline
174.3 & 109.9 & 171.3 &  25.7 & (c) \\
      & 113.4 &   7.4 & 150.5 & (b) \\
      & 159.4 &   6.1 & 145.1 & (e) \\
      & 106.6 & 158.1 &  10.3 & (c) \\
      & 120.3 &  36.9 & 174.3 & (a) \\ \hline \hline
\end{tabular}
\end{center}
\end{table}
\bigskip

The correct solutions in Tables II -- IV are labeled (a).  Solutions with
$\beta > \pi/4$ [labeled (b)] imply $\rho > 1/2$, which is disfavored by the
constraint \cite{DGR} $(\rho^2 + \eta^2)^{1/2} = |V_{ub}/V_{cd}V_{cb}| = 0.27
\pm 0.09$.  Solutions (c) with $\alpha + \gamma > \pi$ similarly imply CKM
parameters outside the currently allowed range of $(\rho,\eta)$, as do those
(e) with $\beta$ or $\gamma$ too small.  A few of the unphysical solutions show
up on the plots of Ref.~\cite{DGR}, but many were not found because attention
was restricted to values of $\alpha$ and $\gamma$ in rough accord with known
CKM constraints. 

One source of discrete ambiguity is the approximate symmetry $\alpha
\leftrightarrow \pi/2 - \alpha$ or $\alpha \leftrightarrow 3 \pi/2 - \alpha$.
This substitution leaves $B,~D,~E$, $F$, and the first term in $C$ unchanged for
fixed values of $\gamma$, $\delta$, $\cat$, $\cap$, and $\tcapp$.  The
substitution does affect the interference terms between $\cat$ and $\cap$ in
$A$ and $C$, but small changes in the parameters seem to be able to compensate
for this effect. 

Another frequently encountered discrete ambiguity involves the interchange
$\gamma \leftrightarrow \delta$, which leaves $D$ and $E$ invariant.  Of
course, $\alpha$ changes under this replacement. 

Many solutions thus can be rejected as unphysical.  Those ``wrong'' solutions
which remain [labeled (d)] are sources of potential ambiguity.  While the
existence of discrete ambiguities undercuts the ability of the method to point
toward new physics, the procedure serves as a consistency check of the standard
CKM picture and as a potential source of further constraints on parameters {\it
within that context}. 
\bigskip

\centerline{\bf ACKNOWLEDGMENTS}
\bigskip

We thank M. Gronau and L. Wolfenstein for fruitful discussions and helpful
suggestions. J. L. R. wishes to acknowledge the hospitality of the CERN and
DESY theory groups during parts of this investigation. This work was supported
in part by the United States Department of Energy under Contract No. DE FG02
90ER40560. 
\bigskip

\def \ajp#1#2#3{Am.~J.~Phys.~{\bf#1}, #2 (#3)}
\def \apny#1#2#3{Ann.~Phys.~(N.Y.) {\bf#1} #2 (#3)}
\def \app#1#2#3{Acta Phys.~Polonica {\bf#1} #2 (#3)}
\def \arnps#1#2#3{Ann.~Rev.~Nucl.~Part.~Sci.~{\bf#1} #2 (#3)}
\def \baps#1#2#3{Bull.~Am.~Phys.~Soc. {\bf#1} #2 (#3)}
\def \cmp#1#2#3{Commun.~Math.~Phys.~{\bf#1} #2 (#3)}
\def \cmts#1#2#3{Comments on Nucl.~Part.~Phys.~{\bf#1} #2 (#3)}
\def \cn{Collaboration}
\def \corn93{{\it Lepton and Photon Interactions:  XVI International Symposium,
Ithaca, NY August 1993}, AIP Conference Proceedings No.~302, ed.~by P. Drell
and D. Rubin (AIP, New York, 1994)}
\def \cp89{{\it CP Violation,} edited by C. Jarlskog (World Scientific,
Singapore, 1989)}
\def \dpff{{\it The Fermilab Meeting -- DPF 92} (7th Meeting of the American
Physical Society Division of Particles and Fields), 10--14 November 1992,
ed. by C. H. Albright \ite~(World Scientific, Singapore, 1993)}
\def \dpf94{DPF 94 Meeting, Albuquerque, NM, Aug.~2--6, 1994}
\def \efi{Enrico Fermi Institute Report No. EFI}
\def \el#1#2#3{Europhys.~Lett.~{\bf#1} #2 (#3)}
\def \f79{{\it Proceedings of the 1979 International Symposium on Lepton and
Photon Interactions at High Energies,} Fermilab, August 23-29, 1979, ed.~by
T. B. W. Kirk and H. D. I. Abarbanel (Fermi National Accelerator Laboratory,
Batavia, IL, 1979}
\def \hb87{{\it Proceeding of the 1987 International Symposium on Lepton and
Photon Interactions at High Energies,} Hamburg, 1987, ed.~by W. Bartel
and R. R\"uckl (Nucl. Phys. B, Proc. Suppl., vol. 3) (North-Holland,
Amsterdam, 1988)}
\def \ib{{\it ibid.}~}
\def \ibj#1#2#3{~{\bf#1}, #2 (#3)}
\def \ichep72{{\it Proceedings of the XVI International Conference on High
Energy Physics}, Chicago and Batavia, Illinois, Sept. 6--13, 1972,
edited by J. D. Jackson, A. Roberts, and R. Donaldson (Fermilab, Batavia,
IL, 1972)}
\def \ijmpa#1#2#3{Int.~J.~Mod.~Phys.~A {\bf#1} #2 (#3)}
\def \ite{{\it et al.}}
\def \jmp#1#2#3{J.~Math.~Phys.~{\bf#1} #2 (#3)}
\def \jpg#1#2#3{J.~Phys.~G {\bf#1} #2 (#3)}
\def \lkl87{{\it Selected Topics in Electroweak Interactions} (Proceedings of
the Second Lake Louise Institute on New Frontiers in Particle Physics, 15--21
February, 1987), edited by J. M. Cameron \ite~(World Scientific, Singapore,
1987)}
\def \ky85{{\it Proceedings of the International Symposium on Lepton and
Photon Interactions at High Energy,} Kyoto, Aug.~19-24, 1985, edited by M.
Konuma and K. Takahashi (Kyoto Univ., Kyoto, 1985)}
\def \mpla#1#2#3{Mod.~Phys.~Lett.~A {\bf#1} #2 (#3)}
\def \nc#1#2#3{Nuovo Cim.~{\bf#1} #2 (#3)}
\def \np#1#2#3{Nucl.~Phys.~B{\bf#1}, #2 (#3)}
\def \pisma#1#2#3#4{Pis'ma Zh.~Eksp.~Teor.~Fiz.~{\bf#1} #2 (#3) [JETP Lett.
{\bf#1} #4 (#3)]}
\def \pl#1#2#3{Phys.~Lett.~{\bf#1}, #2 (#3)}
\def \plb#1#2#3{Phys.~Lett.~B{\bf#1}, #2 (#3)}
\def \pra#1#2#3{Phys.~Rev.~A {\bf#1} #2 (#3)}
\def \prd#1#2#3{Phys.~Rev.~D {\bf#1}, #2 (#3)}
\def \prl#1#2#3{Phys.~Rev.~Lett.~{\bf#1}, #2 (#3)}
\def \prp#1#2#3{Phys.~Rep.~{\bf#1} #2 (#3)}
\def \ptp#1#2#3{Prog.~Theor.~Phys.~{\bf#1}, #2 (#3)}
\def \rmp#1#2#3{Rev.~Mod.~Phys.~{\bf#1} #2 (#3)}
\def \rp#1{~~~~~\ldots\ldots{\rm rp~}{#1}~~~~~}
\def \si90{25th International Conference on High Energy Physics, Singapore,
Aug. 2-8, 1990}
\def \slc87{{\it Proceedings of the Salt Lake City Meeting} (Division of
Particles and Fields, American Physical Society, Salt Lake City, Utah, 1987),
ed.~by C. DeTar and J. S. Ball (World Scientific, Singapore, 1987)}
\def \slac89{{\it Proceedings of the XIVth International Symposium on
Lepton and Photon Interactions,} Stanford, California, 1989, edited by M.
Riordan (World Scientific, Singapore, 1990)}
\def \smass82{{\it Proceedings of the 1982 DPF Summer Study on Elementary
Particle Physics and Future Facilities}, Snowmass, Colorado, edited by R.
Donaldson, R. Gustafson, and F. Paige (World Scientific, Singapore, 1982)}
\def \smass90{{\it Research Directions for the Decade} (Proceedings of the
1990 Summer Study on High Energy Physics, June 25 -- July 13, Snowmass,
Colorado), edited by E. L. Berger (World Scientific, Singapore, 1992)}
\def \stone{{\it B Decays}, edited by S. Stone (World Scientific, Singapore,
1994)}
\def \tasi90{{\it Testing the Standard Model} (Proceedings of the 1990
Theoretical Advanced Study Institute in Elementary Particle Physics, Boulder,
Colorado, 3--27 June, 1990), edited by M. Cveti\v{c} and P. Langacker
(World Scientific, Singapore, 1991)}
\def \yaf#1#2#3#4{Yad.~Fiz.~{\bf#1} #2 (#3) [Sov.~J.~Nucl.~Phys.~{\bf #1} #4
(#3)]}
\def \zhetf#1#2#3#4#5#6{Zh.~Eksp.~Teor.~Fiz.~{\bf #1} #2 (#3) [Sov.~Phys. -
JETP {\bf #4} #5 (#6)]}
\def \zpc#1#2#3{Z.~Phys.~C {\bf#1}, #2  (#3)}

\end{document}